\documentclass[aps,prb,showpacs, twocolumn,amsfonts,amssymb,amsmath,superscriptaddress]{revtex4-1}
\usepackage{graphicx} 
\usepackage{array}
\usepackage{SIunits}
\usepackage{mathtools}
\usepackage{color}
\usepackage{natbib} 
\usepackage{multirow}
\usepackage{amsmath}
\usepackage[colorlinks=true,linkcolor=blue,citecolor=blue]{hyperref}
\newcommand{\pr}{Pr$^{3+}$}

\newcommand{\thf}{$^3$H$_{4}$}

\newcommand{\odt}{$^1$D$_{2}$}

\newcommand{\er}{Er$^{3+}$}

\newcommand{\ho}{Ho$^{3+}$}

\newcommand{\eu}{Eu$^{3+}$}

\newcommand{\YLF}{LiYF$_4$}

\newcommand{\LAWO}{La$_2$(WO$_4$)$_3$}

\newcommand{\wnm}{cm$^{-1}$}

\newcommand{\YSO}[0]{Y$_2$SiO$_5\,$}
\newcommand{\YO}[0]{Y$_2$O$_3\,$}

\newcommand{\DS}{\textcolor{black}} 

\begin{document}

\title{Coherent optical and spin spectroscopy of nanoscale Pr$^{3+}$:Y$_2$O$_3$\\}

\author{D. Serrano}
\affiliation{Chimie ParisTech, PSL University, CNRS, Institut de Recherche de Chimie Paris, F-75005 Paris, France}
\author{C. Deshmukh}
\affiliation{Institut de Ciencies Fotoniques (ICFO), The Barcelona Institute of Science and Technology, 08860 Castelldefels, Barcelona, Spain.}
\author{S. Liu}
\affiliation{Chimie ParisTech, PSL University, CNRS, Institut de Recherche de Chimie Paris, F-75005  Paris,  France}
\author{A. Tallaire}
\affiliation{Chimie ParisTech, PSL University, CNRS, Institut de Recherche de Chimie Paris, F-75005  Paris,  France}
\author{A. Ferrier}
\affiliation{Chimie ParisTech, PSL University, CNRS, Institut de Recherche de Chimie Paris, F-75005  Paris,  France}
\affiliation{Sorbonne Universit\'e, Facult\'e des Sciences et Ing\'{e}nierie, UFR 933, F-75005 Paris, France}
\author{H. de Riedmatten}
\affiliation{Institut de Ciencies Fotoniques (ICFO), The Barcelona Institute of Science and Technology, 08860 Castelldefels, Barcelona, Spain.}
\affiliation{ICREA-Instituci\'{o} Catalana de Recerca i Estudis Avan\c{c}ats, 08015 Barcelona, Spain}
\author{P. Goldner}
\affiliation{Chimie ParisTech, PSL University, CNRS, Institut de Recherche de Chimie Paris, F-75005  Paris,  France}

\date{\today}

\begin{abstract}
We investigate the potential for optical quantum technologies of Pr$^{3+}$:Y$_2$O$_3$ in the form of monodisperse spherical nanoparticles. We measured optical inhomogeneous lines of 27 GHz, and optical homogeneous linewidths of 108 kHz and 315 kHz in particles of 400 nm and 150 nm average diameters respectively for the \odt(0)$\leftrightarrow$\thf(0) transition at 1.4 K. Furthermore, ground state and \odt{} excited state hyperfine structures in \YO are here for the first time determined by spectral hole burning and modeled by complete Hamiltonian calculations. Ground-state spin transitions have energies of 5.99 MHz and 10.42 MHz for which we demonstrate spin inhomogeneous linewidths of 42 and 45 kHz respectively. Spin $T_2$ up to 880 $\mu$s was obtained for the $\pm3/2\leftrightarrow\pm5/2$ transition at 10.42 MHz, a value which exceeds that of bulk Pr$^{3+}$ doped crystals so far reported. These promising results confirm nanoscale Pr$^{3+}$:\YO{} as a very appealing candidate to integrate quantum devices. \DS{In particular, we discuss here the possibility of using this material for realizing spin photon interfaces emitting indistinguishable single photons}.

\end{abstract}

\pacs{42.50.Md,76.30.Kg,76.70.Hb,78.67.$-$n}

\maketitle

\section{Introduction}

Solid-state spins are extensively investigated for quantum-state storage, quantum computation and quantum communications [\onlinecite{awschalom_quantum_2018,PhysRevLett.119.223602,saeedi_room-temperature_2013,Simon2010}]. Among them, rare-earth (RE) spins stand out for presenting electron and/or nuclear spin states with outstandingly long coherence lifetimes ($T_2$) at cryogenic temperatures [\onlinecite{zhong_optically_2015,heinze,rancic_coherence_2017}]. Moreover, RE spins are optically accessible through coherent optical transitions [\onlinecite{thiel_optical_2012}], a unique feature in the solid state which makes them optimum candidates for quantum spin-photon interfaces  [\onlinecite{williamson_magneto-optic_2014}]. In the last years, the potential of RE materials for quantum technologies has been supported by achievements such as high-efficiency optical quantum memories [\onlinecite{hedges_efficient_2010}], quantum teleportation [\onlinecite{bussieres_quantum_2014}], multimode spin-wave storage [\onlinecite{gundogan_solid_2015}] and high-fidelity quantum-state tomography [\onlinecite{rippe_experimental_2008}]. However, an important challenge in bulk crystals remains the addressing of single RE ions, a key point for the development of scalable quantum architectures [\onlinecite{wesenberg_scalable_2007,zhong_nanophotonic_2017,PhysRevLett.120.243601}]. This is due, on the one hand, to the long population lifetimes of the 4$f$-4$f$ optical transitions, and, on the other hand, to the difficulty to isolate a single RE ion within a macroscopic crystal [\onlinecite{kolesov_optical_2012},\onlinecite{eichhammer_spectroscopic_2015}]. A promising approach to overcome these difficulties consists in coupling RE emitters in nanocrystals to high-quality-factor optical microcavities [\onlinecite{casabone_cavity-enhanced_2018}]. In particular, using nanoscale crystals facilitates reaching the single emitter level [\onlinecite{eichhammer_spectroscopic_2015}] whereas the cavity coupling provides stronger light-matter interactions [\onlinecite{burek_fiber-coupled_2017},\onlinecite{zhong_nanophotonic_2017}] and Purcell-enhanced spontaneous emission rates [\onlinecite{casabone_cavity-enhanced_2018}]. 

Exploiting this approach to build novel quantum devices strongly relies on developping RE nanocystals with narrow optical and spin homogeneous lines. Those are not straightforwardly available as nanomaterials and often show additional homogeneous linewidth broadening, i.e. $T_2$ shortening, due to surface states and modified spin bath dynamics [\onlinecite{knowles_observing_2014,lutz_effects_2017,meltzer_photon_2004}]. Still, very promising results have been recently obtained with chemically synthesized $^{151}$Eu$^{3+}$ doped Y$_2$O$_3$ nanocrystals. 
Y$_2$O$_3$ is a low-magnetic-moment density host which can be obtained with accurate control over particle size and morphology [\onlinecite{de_oliveira_lima_influence_2015,liu_controlled_2018}]. More important, optical homogeneous linewidths down to 25 kHz have been reported in these nanocrystals  for the $^7$F$_0\leftrightarrow^5$D$_0$ transition at 580.883 nm [\onlinecite{bartholomew_optical_2017},\onlinecite{liu_controlled_2018}], and millisecond-long nuclear spin $T_2$ [\onlinecite{serrano_all-optical_2017}]. A disadvantage of Eu$^{3+}$ is however that it presents weak oscillator strength ($\sim$10$^{-8}$ in \YO [\onlinecite{morrison_optical_1983}]) and low emission branching ratio for the $^7$F$_0\leftrightarrow^5$D$_0$ line ($\sim$0.016 [\onlinecite{casabone_cavity-enhanced_2018}]). This reduces the achievable Purcell enhancement by a factor 60 for this ion [\onlinecite{casabone_cavity-enhanced_2018}]. In these aspects, a good alternative rare-earth ion to Eu$^{3+}$ is Pr$^{3+}$. In Y$_2$O$_3$, the Pr$^{3+}$:$^{1}$D$_2(0)\leftrightarrow^3$H$_4(0)$ transition is expected to exhibit, at least, one order of magnitude larger oscillator strength than the Eu$^{3+}$: $^7$F$_0\leftrightarrow^5$D$_0$ transition [\onlinecite{krupke},\onlinecite{morrison_optical_1983}]. Higher emission branching ratio is equally expected from luminescence investigations carried out in this compound [\onlinecite{guyot_luminescence_1996}]. However, very few previous studies exist on optical homogeneous lines in Pr$^{3+}$:Y$_2$O$_3$ [\onlinecite{okuno_homogeneous_1994,okuno_two_1995,okuno_1999}] while hyperfine structures and spin homogeneous lines have never been reported.

In the present work, we carry out a complete high-resolution and coherent optical and nuclear spin spectroscopic investigation of $^{141}$Pr$^{3+}$:\YO ceramics and nanoparticles. The article is organized as follows: experimental details and methods are given in Sec. \ref{II}. In Sec. \ref{A}, optical inhomogeneous and homogeneous linewidths results are shown and discussed. We next present the experimental investigation of the $^{141}$\pr hyperfine structures in \YO (Sec. \ref{B}). This is followed by hyperfine structure and g-factor calculations in Sec. \ref{C}. In Sec. \ref{D}, spin resonance results are reported from which we derive ground state spin inhomogeneous linewidths and coherence lifetimes. Finally, a summary of results and discussion about the potential of \pr:\YO for quantum technologies applications is given in Sec. \ref{E}.

\section{Experimental}\label{II}

Spherical monodispersed 0.05\%Pr$^{3+}$:Y$_2$O$_3$ nanoparticles were synthesized by homogeneous precipitation with average particle sizes of 400 and 150 nm, and crystalline domains of 120 nm and 80 nm respectively as determined by X-ray diffraction [\onlinecite{de_oliveira_lima_influence_2015}]. Post-synthesis, the particles were placed onto a glass plate and submitted twice to a pure oxygen plasma for 3 minutes. A home-made system operating at a frequency of 2.45 GHz, a microwave power of 900 W and a pressure of 1 mbar was used [\onlinecite{liu_controlled_2018},\onlinecite{Shuping_mw}]. A Pr$^{3+}$:Y$_2$O$_3$ ceramic (0.05\% at.) was elaborated by mixing stoichiometric Y$_2$O$_3$ and Pr$_6$O$_{11}$ oxides (99,99\% purity) and by pressing these powders into pellets followed by sintering at 1500 $^\circ$C for 48 h. Particles and ceramics were obtained with pure cubic phase (Ia-3 space group), where Pr$^{3+}$ ions replace Y$^{3+}$ cations at $C_{3i}$ and $C_2$ point symmetry sites [\onlinecite{okuno_1999}]. In the following, we will focus on the spectroscopic properties of \pr{} ions occupying $C_2$ sites.

\begin{figure}
	\centering
	\includegraphics [width=3.5 in]{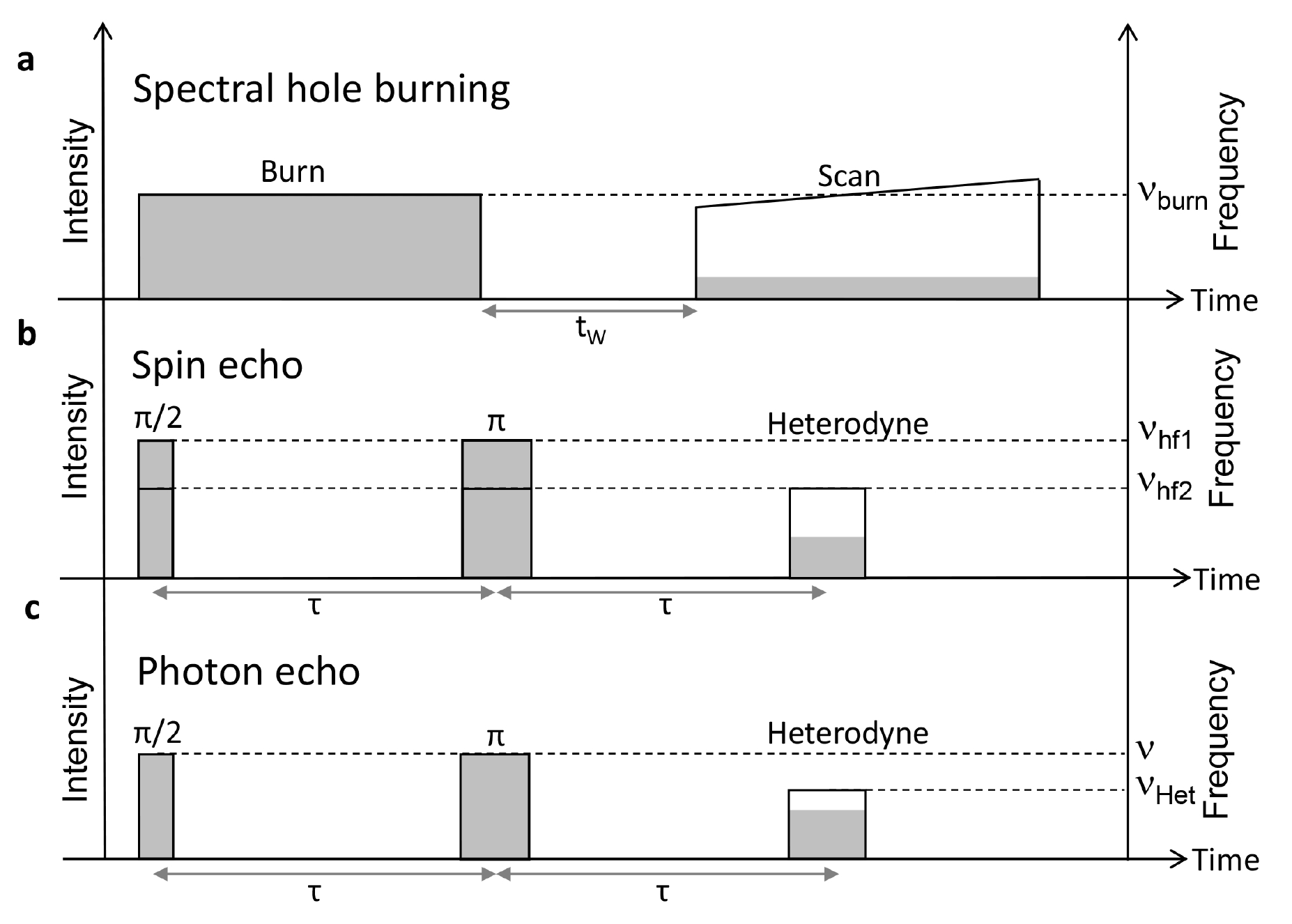}
	\caption{Pulse sequences. (a) Spectral hole burning sequence. (b) Two-pulse Raman spin echo sequence. The spin-echo amplitude at 2$\tau$ is probed by a single frequency pulse referred to as heterodyne pulse. (c) Two-pulse photon echo sequence with heterodyne detection. We note that the role of the heterodyne pulse is not the same in (b) and (c). In (c), it is used as a local oscillator while in (b) it also converts spin coherence into an optical one for detection. Thus, in the latter case, its frequency must match one of the two colors in the bi-frequency pulses (either $\mathrm{\nu_{hf1}}$ or $\nu_\mathrm{hf2}$). Typical pulse lengths are 100s of $\mu$s in (a), 10s of $\mu$s in (b) and 100s of ns in (c). Gray areas represent pulse intensities, and black lines pulse frequencies.}
	\label{fig.1}
\end{figure}

Low-temperature high-resolution and coherent spectroscopy investigations were carried out on ensembles of nanoparticles in the form of powders. Those were placed between two glass plates forming a layer of about 500 $\mu$m thick. The ceramic sample was cut into slices of 250 $\mu$m for measurements. Cryogenic conditions were provided by a He bath cryostat (Janis SVT-200), operated in liquid mode for T $<$ 2 K, and gas flow mode for T $>$ 2 K. The temperature was monitored directly on the sample holders by a Si diode (Lakeshore DT-670). Excitation was provided by a CW dye laser (Sirah Matisse DS) with $\sim$200 kHz linewidth. Pulsed sequences required for spectral hole burning, spin echo and photon echo measurements (Fig. \ref{fig.1}) were created by modulating the CW laser output with two acousto-optic modulators (AOMs) driven by an arbitrary waveform generator (Agilent N8242A). For more details, refer to the optical setup scheme and description in reference [\onlinecite{serrano_all-optical_2017}]. Light emerging from the backside of the samples was collected by a series of lenses, as proposed in [\onlinecite{perrot}] for highly-scattering media, and focused on an avalanche photodiode (Thorlabs A/M110) for detection. Photoluminescence (PL) spectra were recorded at 10 K by exciting ensembles of nanoparticles \DS{at 460 nm (\thf$\rightarrow$$^3$P$_1$,$^1$I$_6$)} with an optical parametric oscillator (OPO) pumped by the third harmonic (355 nm) of a Nd:YAG laser (Ekspla NT342B). Emitted light was sent to a spectrometer (Acton SP2300) and detected by a cooled down CCD camera (Princeton Instruments). The \pr:\YO absorption spectrum was acquired at 15 K from a 28-mm-thick transparent ceramic, using a Varian Cary 6000i spectrophotometer. The integrated absorption coefficient for the \thf(0)$\rightarrow$\odt(0) transition enabled calculating the transition oscillator strength from which we determined the transition branching ratio (see Sec. \ref{E}). An optical closed-cycle cryostat was used in photolumiescence and absorption measurements.

Optical inhomogeneous lines ($\Gamma_{\mathrm{inh}}$) were measured at 12 K for the $^3$H$_{4} (0)\rightarrow^1$D$_{2} (0)$ optical transition, where ``0'' refers here to the lowest energy levels in the $^3$H$_{4}$ and $^1$D$_{2}$ electronic multiplets. The line profile was obtained by photoluminescence excitation (PLE), by collecting $^1$D$_{2}$ emissions while scanning the CW laser wavelength around the peak maximum at $\lambda_{vac}$ = 619.011 nm. A long-pass filter was placed in front of the detector to filter out excitation light. Persistent spectral holes were burned at 1.4 K by a single 500-$\mu$s-long pulse, and then probed at time $t_W$, by monitoring the transmission intensity of a weak scanning pulse (Fig. \ref{fig.1}(a)). The waiting time, $t_W$, is set so that $t_W>\mathrm{\tau_{1D2}}$, with $\mathrm{\tau_{1D2}}$ the \odt$ $ excited state population lifetime or optical T$_1$. Spin inhomogeneous and homogeneous lines were obtained at 1.4 K from two-pulse Raman spin echoes (Fig. \ref{fig.1}(b)). A spin population difference was first induced by optically pumping a sub-ensemble of ions from one spin level to the other one [\onlinecite{serrano_all-optical_2017}]. Then, two-color laser pulses were applied to create and rephase spin coherences. The spin inhomogeneous line was measured by varying the frequency detuning between the two colors ($\nu_\mathrm{hf1}$ - $\nu_\mathrm{hf2}$) while keeping a fixed pulse delay $\tau$ [\onlinecite{serrano_all-optical_2017}]. Spin $T_2$ values were directly derived from the decay of the spin echo amplitude when increasing $\tau$. Finally, optical coherence lifetimes were measured for the \thf(0)$\leftrightarrow$\odt(0) transition at 1.4 K by two-pulse photon echo spectroscopy with heterodyne detection [\onlinecite{perrot}] (Fig. \ref{fig.1}(c)). $\pi$-pulse lengths were set to 15 $\mu$s in Fig. \ref{fig.1}(b) and 700 ns in Fig. \ref{fig.1}(c), with an input power of approximately 50 mW.

\section{Results}\label{III}

\subsection{Optical inhomogeneous and homogeneous linewidths}\label{A}

\begin{figure}
	\centering
	\includegraphics [width=3.5 in]{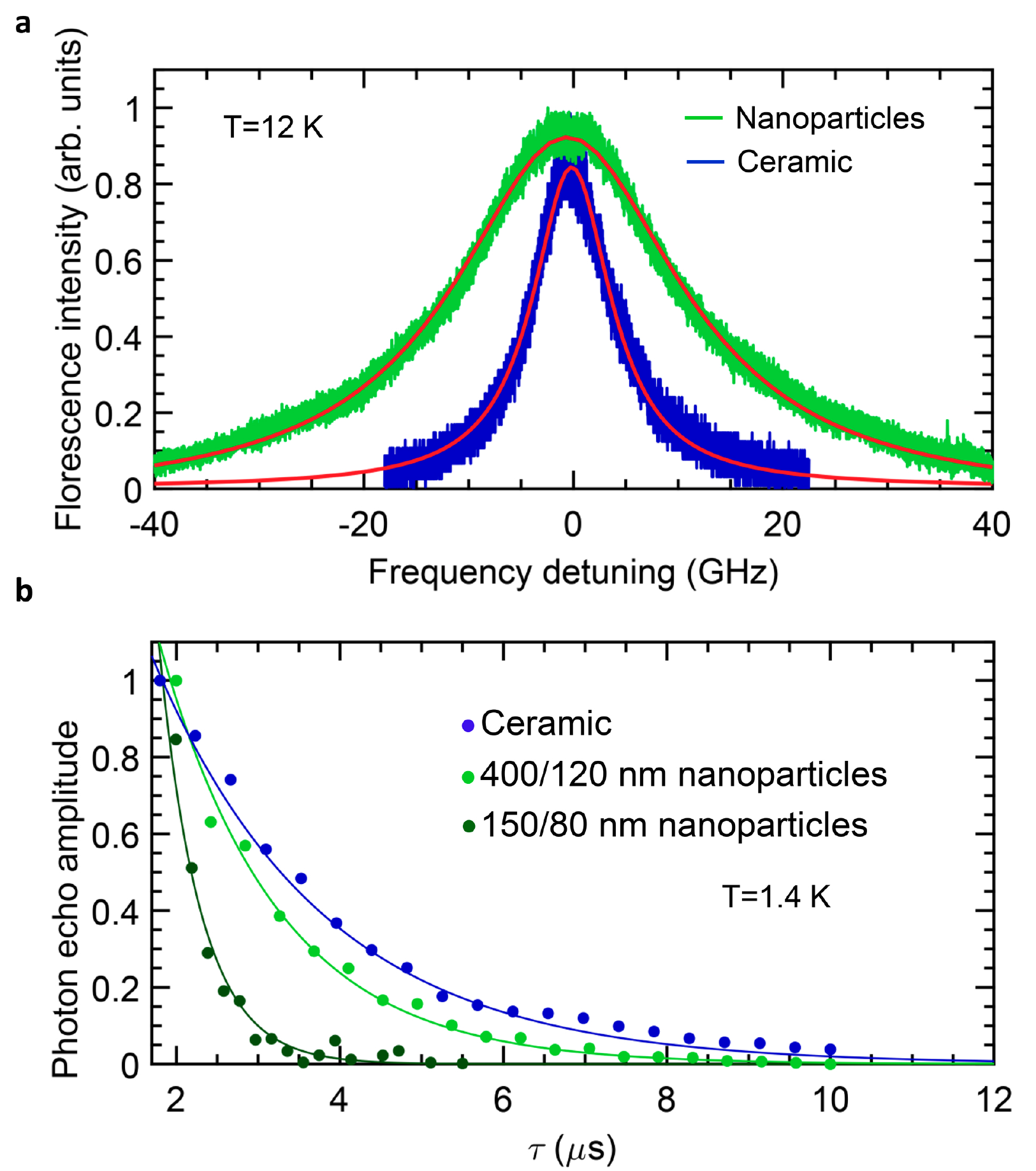}
	\caption{Optical inhomogeneous lines and coherence lifetimes. (a) Optical inhomogeneous lines for the \thf(0)$\leftrightarrow$\odt(0) transition for ceramic (blue line) and 400/120 nm diameter particles/crystallites (green line). Red lines correspond to Lorentzian fits. Zero frequency detuning stands for 484.308 THz for the ceramic and 484.317 THz for nanoparticles. (b) Photon echo decays from ceramic (blue), 400/120 nm diameter particles/crystallites (green) and 150/80 nm diameter particles/crystallites (dark green). All decays show single exponential character, therefore, optical $T_{2,}$values of 4.5$\pm$0.5 $\mu$s, 3.0$\pm$0.3 $\mu$s and 1.0$\pm$0.1 $\mu$s were obtained by fitting the photon echo amplitudes to $\exp(-2\tau/T_{2})$ (lines). The resulting homogeneous lines correspond to 72$\pm$16 kHz, 108$\pm$21 kHz and 315$\pm$64 kHz for ceramic, 400-nm and 150-nm particles respectively.}
	\label{fig.2}
\end{figure}

$^3$H$_{4}(0)\leftrightarrow^1$D$_{2}(0)$ optical inhomogeneous lines are displayed in Fig. \ref{fig.2}(a) for the ceramic and 400-nm-diameter nanoparticles. The ceramic sample shows a full-width half maximum (FWHM) of 9 GHz and has Lorentzian profile. This is consistent with previous investigations on bulk \YO:Pr$^{3+}$ [\onlinecite{okuno_1999}] and \YSO:Pr$^{3+}$ [\onlinecite{equall_homogeneous_1995}]. In opposition, a three times broader line i.e. 27 GHz, was obtained in the nanoparticles case. This line broadening is due to the post-synthesis O$_2$ plasma treatment. Indeed, similar behaviour is observed in Eu$^{3+}$:\YO nanoparticles: bulk-like inhomogeneous lines are measured after synthesis [\onlinecite{bartholomew_optical_2017}] whereas a factor 2-3 line broadening is observed following  O$_2$ plasma processing [\onlinecite{liu_controlled_2018}]. The processing also provokes a blueshift of about 10 GHz from 484.308 THz ($\lambda_{vac}$=619.011 nm), as measured in the ceramic, to 484.317 THz. Shift and broadening are attributed to strain induced by the increased oxygen content in the nanoparticles under $O_2$ plasma treatment [\onlinecite{Shuping_mw}]. 

While the plasma processing broadens the optical inhomogeneous line, it has positive impact on the Pr$^{3+}$ optical coherence lifetime. Indeed, longer coherence lifetimes are found after O$_2$ plasma processing. In particular, coherence lifetimes of 3 $\mu$s and 1 $\mu$s (Fig. \ref{fig.2}(b)), corresponding to homogeneous lines $\Gamma_{\mathrm{h}}$=$(\pi T_2)^{-1}$ of 108 kHz and 315 kHz were obtained in 400-nm and 150-nm-diameter particles respectively. This is about a factor of 2 less than homogeneous linewidths obtained immediately after synthesis. Moreover, these values are narrower than those reported in some particular bulk \pr:Y$_2$O$_3$ crystals [\onlinecite{okuno_1999}] although they still remain far from the 1.1 kHz linewidth given by the optical $T_1$ limit. $T_1$ is found equal to 140 $\mu$s in the nanoparticles, which is similar to bulk \pr:Y$_2$O$_3$ [\onlinecite{guyot_luminescence_1996}]. The decay curve is later shown in Section \ref{E}. 

In a previous work, the dominant optical dephasing mechanism in \eu:\YO nanoparticles has been attributed to fluctuating electric fields associated to charged surface states [\onlinecite{bartholomew_optical_2017}]. The broader homogeneous line found here for the 150-nm-diameter particles (Fig. \ref{fig.2}(b)) is consistent with this hypothesis since surface to volume ratios increase as the particle size decreases [\onlinecite{bartholomew_optical_2017}]. Nonetheless, it seems that the observed broadening is better explained by the decrease in crystalline grain size rather than particle size. Crystalline domains decrease from 120 nm to 80 nm from 400-nm-diameter to 150-nm-diameter particles. Therefore, the electric field strength ($\mathbf{E}\propto r^{-2}$) at the center of a crystallite is about 2.3 times larger in an 80-nm crystallite than a 120-nm one. This assumes that the field originates from electric charges located at the interface between crystallites. In contrast, if we take into account electric noise originated at the nanoparticles outer surface, 7 times larger electric field strength is expected in the 150-nm particles. This should lead to larger broadening than experimentally observed. In a similar way, the impact of the crystalline domain size over the particle size in the optical dephasing has been recently evidenced on Eu$^{3+}$\YO nanoparticles [\onlinecite{liu_controlled_2018}]. 

\begin{center}
\subsection{Spectral hole burning}\label{B}
\end{center}

\begin{figure}
	\centering
	\includegraphics [width=3.5 in]{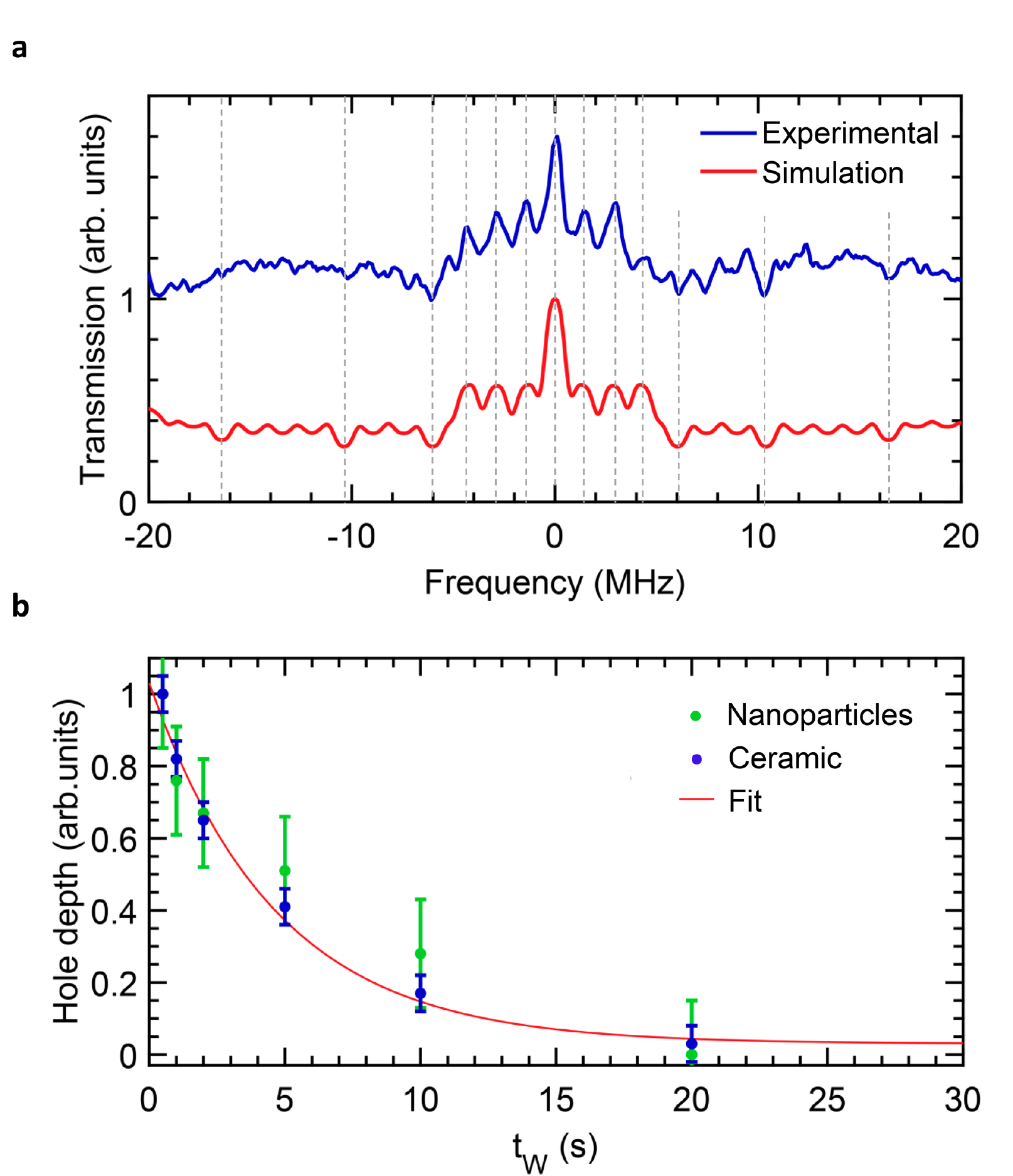}
	\caption{Hyperfine level structure and spin population lifetime (a) Single-shot spectral hole burning spectrum from Pr$^{3+}$:\YO ceramic (blue) compared to simulation (red). Main transitions are indicated by dashed lines. (b) Hole decay as a function of waiting time before readout ($t_W$) as measured for ceramic (green dots) and 400/120 nm particles (blue dots). Single exponential fit to the ceramic data reveals a spin relaxation time of  5 $\pm$ 1 s (red line).}
	\label{fig.3}
\end{figure}

\begin{figure}
	\centering
	\includegraphics [width=2.2 in]{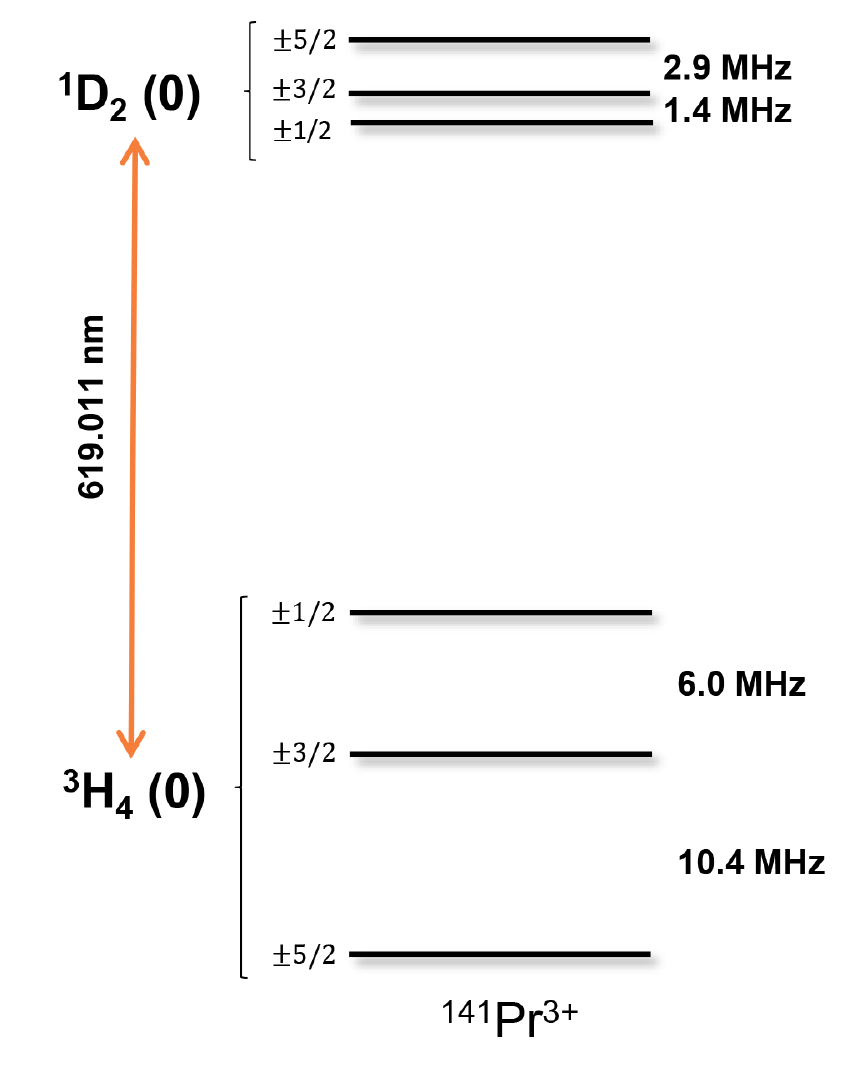}
	\caption{Measured hyperfine structure in $^{141}$Pr:\YO{} ($I=5/2$). $\pm M_I$ quantum numbers are only used as labels and do not represent actual eigenstate projections.}
	\label{fig.4}
\end{figure}

With a single nuclear isotope and nuclear spin $I=$ 5/2, the $^{141}$Pr$^{3+}$ hyperfine structure at zero magnetic field consists of three doubly-degenerated levels, namely $\pm$1/2, $\pm$3/2 and $\pm$5/2 [\onlinecite{equall_homogeneous_1995}]. The hyperfine energy splittings in Y$_2$O$_3$ were obtained for the $^3$H$_4$(0) ground state and $^1$D$_2$(0) excited state energy levels by spectral hole burning (SHB). The recorded SHB spectrum is displayed in Fig. \ref{fig.3}(a), with the main hole appearing at the center surrounded by side holes and anti-holes. We derived the excited state splittings from the side holes position [\onlinecite{Goldner:2015ve}], finding that it corresponds to 2.9 MHz and 1.4 MHz. The anti-holes at 6 MHz, 10.4 MHz and 16.4 MHz reveal the ground-state $\pm1/2\leftrightarrow\pm3/2$, $\pm3/2\leftrightarrow\pm5/2$ and $\pm1/2\leftrightarrow\pm5/2$ hyperfine splittings respectively. A simulated spectrum computed with the mentioned ground and excited state hyperfine splittings shows very good agreement with the experiment. More accurate values for the ground state hyperfine splittings, up to 10 kHz precision, are given in Section \ref{D} derived from spin resonance investigations. Ground and excited state energy schemes are shown in Fig. \ref{fig.4}.\\ 

The central hole decay as a function of time is shown in Fig. \ref{fig.3}(b). The hole lifetime is estimated to 5 $\pm$ 1 s  by single exponential fit to the ceramic data. A single exponential fit results in a longer decay time for the nanoparticles (7 $\pm$ 3 s). Nevertheless, the difference is not considered significant as it falls within experimental error bars. Those are quite important, especially in the nanoparticles case, due to the strong light scattering at the samples resulting in rather low signal to noise ratios. The measured hole decay time results from a combination of different population decay rates among the three ground state hyperfine levels [\onlinecite{suter}]. Still, it provides a  good indication of the spin relaxation time ($T_1$). The obtained values are comparable to that reported in other Pr$^{3+}$ doped crystals such as YAlO$_3$ and La$_2$(WO$_4$)$_3$ [\onlinecite{lovric_hyperfine_2011,suter}]. This is however two orders of magnitude lower than values typically reported in \pr:\YSO [\onlinecite{holliday_spectral_1993}] although a shorter $T_1$ component, of 7 s, has been also observed in this crystal [\onlinecite{ETH},\onlinecite{Gundogan}].\\

\subsection{Hyperfine structures  and g-factors calculation}\label{C}
\label{CF}
The energy level structure of rare-earth ions is the result of several interactions which can be listed as:
\begin{equation}
H=[H_{FI}+H_{CF}]+[H_{HF}+H_{Q}+H_{Z}+H_{z}]\label{eq1}
\end{equation}

The first bracketed term, including the free-ion Hamiltonian ($H_{FI}$) and the crystal-field Hamiltonian ($H_{CF}$) determines the energies and wavefunctions of the so-called crystal field levels. The second term is formed of the hyperfine interaction ($H_{HF}$), the nuclear quadrupole interaction ($H_{Q}$), and the electron ($H_{Z}$) and nuclear ($H_{z}$) Zeeman interactions. For non-Kramers RE ions, i.e. with an even number of $f$ electrons, electron Zeeman and hyperfine interactions have zero diagonal matrix elements for non-degenerate crystal field (CF) levels. This is the case for \pr{} ions in $C_2$ site symmetry. When such levels are separated from other CF levels by more than a few \wnm{}, electron Zeeman and hyperfine interactions have only second-order contributions to hyperfine structures, resulting in splittings on the order of 10s to 100s of MHz [\onlinecite{Goldner:2015ve}]. 

Previous work showed that experimental rare earth fine and hyperfine structures can be well reproduced by the Hamiltonian of Eq. \ref{eq1}. Some examples are \pr [\onlinecite{lovric_hyperfine_2011},\onlinecite{guillot-noel_calculation_2010}], \er [\onlinecite{Marino:2012ug},\onlinecite{Horvath:2018wt}] and \ho [\onlinecite{Wells:2004ie}] doped crystals. Following our work on \pr{} doped \YLF{} [\onlinecite{Goldner:2015ve}] and \LAWO{} [\onlinecite{guillot-noel_calculation_2010}], hyperfine structures in \YO were here computed by directly diagonalizing the Hamiltonian in Eq. \ref{eq1}. For $H_{FI}$ and $H_{CF}$, we used the parameters determined by Morrison et al. in an extensive study of CF energy levels modeling in RE doped \YO [\onlinecite{morrison_optical_1983}]. Matrix elements of $H_{HF}+H_{Q}+H_{Z}+H_{z}$ were then evaluated using the following equations [\onlinecite{guillot-noel_calculation_2010},\onlinecite{Goldner:2004kn}]:

\begin{eqnarray}
H_{HF}&=&a_{1}H_{1}\label{eq2-1}\\
H_{Q}&=&a_2H_{2}+a_3H_{3} \label{eq2-2}\\
H_Z&=&\mu_B \mathbf{B}\cdot (\mathbf{L}+g_e\mathbf{S})\label{eq2-3}\\
H_z&=&-g_n \mu_n \mathbf{B}\cdot\mathbf{I} \label{eq2-4}
\end{eqnarray}

where $\mu_b$ amd $\mu_n$ are the  Bohr and nuclear magnetons, $\mathbf{B}$ an external magnetic field, $\mathbf{L}$, $\mathbf{S}$ and $\mathbf{I}$, the orbital, electron and nuclear spin angular momenta, $g_e$ the electron $g$-factor and $g_n = 1.6$ is the nuclear $g$-factor of $^{141}$\pr. The definitions of $a_i$ are given in Appendix. These $a_i$, which contain several parameters not precisely known like $4f$ electron radius or screening factors, were adjusted to give the best fit to the \thf(0) and \odt(0) experimental hyperfine splittings (Fig. \ref{fig.4}). The resulting values are $a_1 = 660$ MHz, $a_2 = 18.6$ MHz and $a_3 = 4.7 \times 10^{-8}$. Note that $a_3$, which corresponds to the lattice contribution to the quadrupole interaction is dimensionless, whereas rank 2 CF parameters appear in $H_3$. These values are reasonably close to those determined in previous studies. For example, in \pr:\LAWO,  $a_1 = 721$ MHz, $a_2 = 26.9$ MHz and $a_3 = 1.7 \times 10^{-7}$ [\onlinecite{guillot-noel_calculation_2010}]. 

We first discuss the zero field case, i.e. $\mathbf{B}=0$ in Eqs. \ref{eq2-3}-\ref{eq2-4}. The calculated and experimental splittings, in good agreement, are shown in Table \ref{table1}. 
In opposition to \pr:\LAWO, the excited state splittings were correctly calculated using the ${^1}$D$_2$(0) CF level and not ${^1}$D$_2$(1). This suggests a more accurate CF analysis in the case of \pr:\YO{} in which the true site symmetry $C_2$ was used, whereas a $C_{2v}$ symmetry was substituted to $C_1$ in \pr:\LAWO{} [\onlinecite{guillot-noel_calculation_2010}]. Diagonalizing the total Hamiltonian of Eq. \ref{eq1} when setting the quadrupolar interaction $H_{Q}$ to zero shows that ground state splittings are dominated by the hyperfine interaction, which gives more than 90 \% of the observed values. In the excited state, however, it has a contribution lower than $ 0.2$ MHz. This is consistent with the hyperfine interaction being a  second-order perturbation: the ground state first CF level energies (0, 108, 291, 345 \wnm) are closer than the excited ones (0, 267, 711 \wnm), which qualitatively explains the difference in hyperfine contribution. A more precise assessment would take into account matrix elements of the form $<n|J_i|n'><n'|J_j|n>$, where $n,n'$ denote CF levels of the multiplet of interest and $J_{i,j}$ with $i,j=x,y,z$ the total angular momentum along the  CF axes [\onlinecite{lovric_hyperfine_2011}]. The low hyperfine contribution shows that the excited state splittings are mainly due to the quadrupole interaction. This one has two parts that correspond to the interaction between the nuclear quadrupole moment and the electric field gradient created on one hand by the $f$ electrons and on the other hand by the lattice electrons (respectively $a_2H_{2}$ and $a_3H_{3}$ in Eq. \ref{eq2-2}). Setting them to zero alternatively suggests that they have contributions of opposite signs. This gives small excited state splittings in comparison with \pr{} in \LAWO{} and in site 1 of \YSO{}  (Table \ref{table1}). The same holds for the ground state splittings, which is likely to be explained by the larger CF splittings for \pr{} in \YO{} (0, 108, 291, 345 \wnm) than in \LAWO{} (0, 59, 91,168 \wnm) [\onlinecite{guillot-noel_calculation_2010}]  or \YSO{} (0, 88, 146  \wnm) [\onlinecite{equall_homogeneous_1995}].

Applying an external magnetic field removes the two-fold degeneracy of each hyperfine level, because of the hyperfine and electron Zeeman second-order cross-term, and the nuclear Zeeman one. To determine $^{141}$\pr{} corresponding gyromagnetic factors, we switched  to a spin Hamiltonian approach, in which the hyperfine structure of a given CF level is given by the Hamiltonian:
\begin{equation}
H_{s}=\mathbf{B}\cdot\mathbf{M}\cdot\mathbf{I}+\mathbf{I}\cdot\mathbf{Q}\cdot\mathbf{I},
\end{equation}
where $\mathbf{M}$ and $\mathbf{Q}$ are the effective Zeeman and quadrupolar tensors. 
In their respective principal axes, the $\mathbf{M}$ and $\mathbf{Q}$ tensors read:
\begin{eqnarray}
  \mathbf{M} &=& \left(
    \begin{array}{ccc}
      g_{1}	& 		0 		& 0 \\
      0			& g_{2}		& 0 \\
      0			&				&g_{3}
    \end{array} 
  \right)   \label{eq_M}  \\
  \mathbf{Q} &=&	 \left(
    \begin{array}{ccc}
      E-\frac{1}{3}D 	& 		0 		& 0 \\
      0			& -E-\frac{1}{3}D	& 0 \\
      0			&				& \frac{2}{3}D
    \end{array} 
  \right).	
  \label{eq_Q}
\end{eqnarray}
One principal axis of both tensors coincides with the site $C_2$ symmetry axis. The $g_i$, $D$ and $E$ values were determined by a fit to the calculated hyperfine splittings given by the Hamiltonian of Eq. \ref{eq1} with a fixed magnetic field magnitude of 5 mT and a direction that spanned a sphere in the CF axes system, as in previous experiments [\onlinecite{Longdell:2002bv},\onlinecite{lovric_hyperfine_2011}]. The fitted parameters are given in Table \ref{table2}. The largest ground state gyromagnetic factor is 85 MHz/T, and results mainly from the second-order perturbation as the nuclear Zeeman contribution is only 12.2 MHz/T. It is 
smaller than the values for \LAWO{} (147 MHz/T) [\onlinecite{lovric_hyperfine_2011}] and \YSO{} (113 MHz/T) [\onlinecite{Longdell:2002bv},\onlinecite{Lovric:2012ca}]. This can be again explained by the larger CF splittings in \YO{} that reduce second-order effects. In the excited state, the gyromagnetic factors are close to the nuclear contribution of 12.2 MHz/T, because of lower second-order effects, as already observed above for the zero field splittings.

\begin{table}
\begin{tabular}{r| c c |c| c}
\hline
\hline
        & \multicolumn{2}{ c| }{Y$_2$O$_3$} & La$_2$(WO$_4$)$_3$ & Y$_2$SiO$_5$ \\
       
   		& Exp. (MHz) & Calc. (MHz) & Exp. (MHz) & Exp. (MHz) \\
		\hline
\multirow{2}{*}{$^3$H$_4$} & 6.0 & 5.4 & 14.90 & 10.19 \\
                     & 10.4 & 10.7 & 24.44 & 17.30 \\
        \hline
\multirow{2}{*}{$^1$D$_2$} & 1.4 & 1.5 & 4.94 & 4.59 \\
                     & 2.9 & 2.9 & 7.23 & 4.84 \\
 
  \hline
  \hline
 
\end{tabular}
\caption{Experimental and calculated hyperfine splittings at zero magnetic field for ground  $^3$H$_{4}$(1) and excited $^1$D$_{2}$(1) states for $^{141}$Pr$^{3+}$ in Y$_2$O$_3$ ($C_2$ site), La$_2$(WO$_4$)$_3$ [\onlinecite{lovric_hyperfine_2011}] and Y$_2$SiO$_5$ (site 1) [\onlinecite{equall_homogeneous_1995}].}
\label{table1}
\end{table}

\begin{table}
\begin{tabular}{r| c c}

 \hline
 \hline

   	 Parameter & Ground state & Excited state \\
   	  \hline
   	 D (MHz) & -2.66 & 0.72 \\
	$|E|$ (MHz) & 0.2 & 0.078 \\
	 $|g_1|$ (MHz/T) &   13.9 & 10.8 \\
	 $|g_2|$ (MHz/T) & 18.6 & 12.9 \\
	 $|g_3|$ (MHz/T) & 84.7 & 14.9 \\
 \hline
 \hline
\end{tabular}
\caption{Ground and excited state spin Hamiltonian parameters fitted to calculated hyperfine splittings under a magnetic field of 5 mT which directions follows the surface of a sphere. }
\label{table2}
\end{table}

\subsection{Spin inhomogeneous and homogeneous linewidths}\label{D}

\begin{figure}
	\centering
	\includegraphics [width=3.5 in]{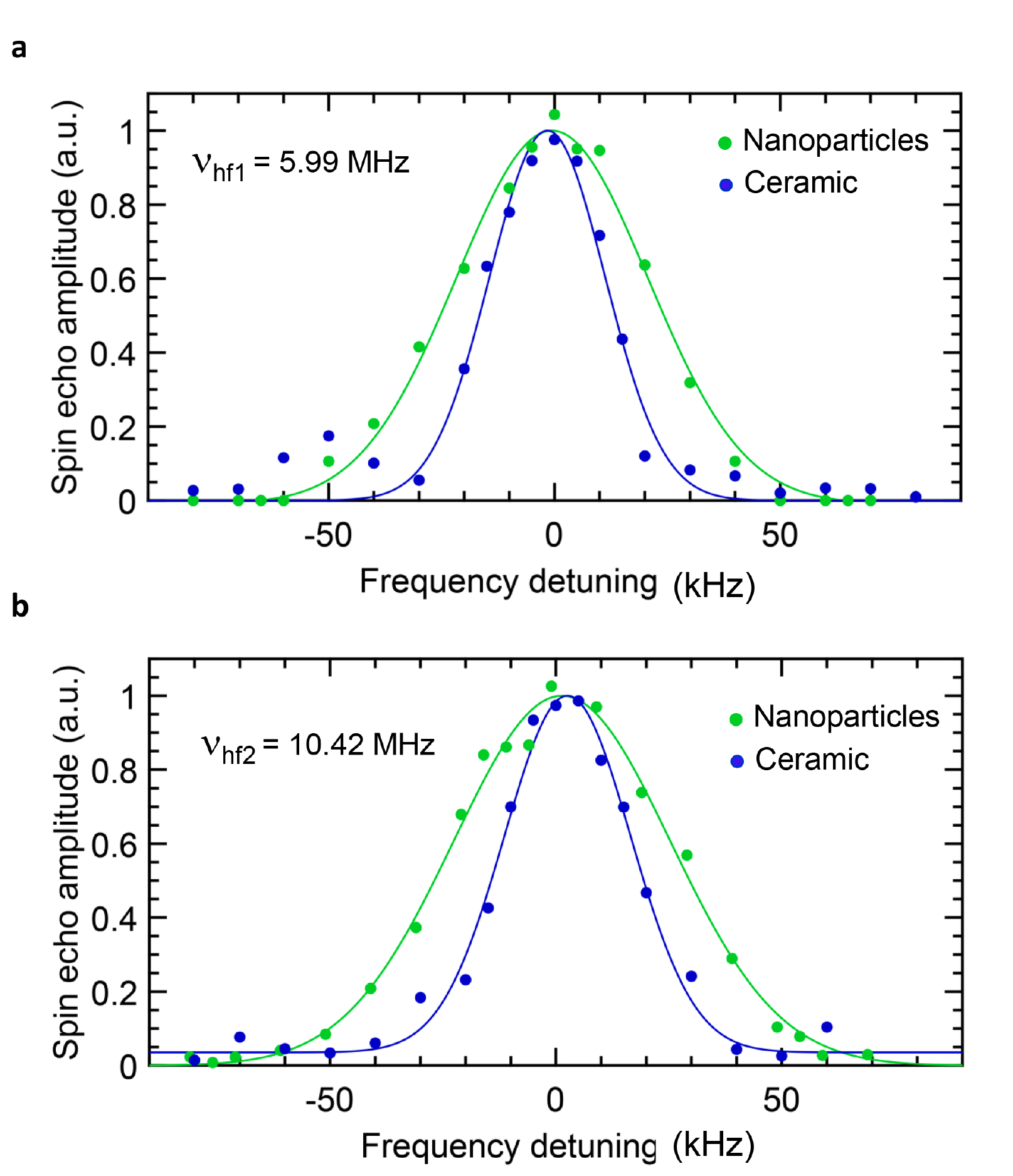}
	\caption{Spin inhomogeneous linewidth obtained by two-pulse Raman spin echo at 1.4 K for ceramic (blue) and 400/120 nm diameter particles/crystallites (green). (a) $\pm1/2\leftrightarrow\pm3/2$ transition at 5.99 MHz. (b) $\pm3/2\leftrightarrow\pm5/2$ transition at 10.42 MHz. Lines correspond to fitted curves to the experimental data (dots).}
	\label{fig.5}
\end{figure}

The evolution of the spin-echo amplitude as a function of the two-color frequency detuning is displayed in Fig. \ref{fig.5}.  Maximum spin echo amplitudes were found for frequency detunings of 5.99 MHz and 10.42 MHz, providing one order of magnitude better accuracy for the ground-state hyperfine splittings than hole burning experiments (Sec. \ref{B}). The analysis gives a linewidth of 29$\pm$2 kHz for both spin transitions in the ceramic sample, compatible with bulk \pr:\YSO values [\onlinecite{gundogan_solid_2015}]. In nanoparticles, larger spin inhomogeneous linewidths of 42$\pm$9 and 48$\pm$6 kHz were measured for the $\pm1/2\leftrightarrow\pm3/2$ and the $\pm3/2\leftrightarrow\pm5/2$ transitions respectively. We believe that this can be also attributed to the O$_2$ plasma processing as non-processed \eu doped nanoparticles present spin inhomogeneous linewidths equivalent to ceramics [\onlinecite{serrano_all-optical_2017}] and bulk crystals [\onlinecite{arcangeli_spectroscopy_2014}]. 

\begin{figure}
	\centering
	\includegraphics [width=3.5 in]{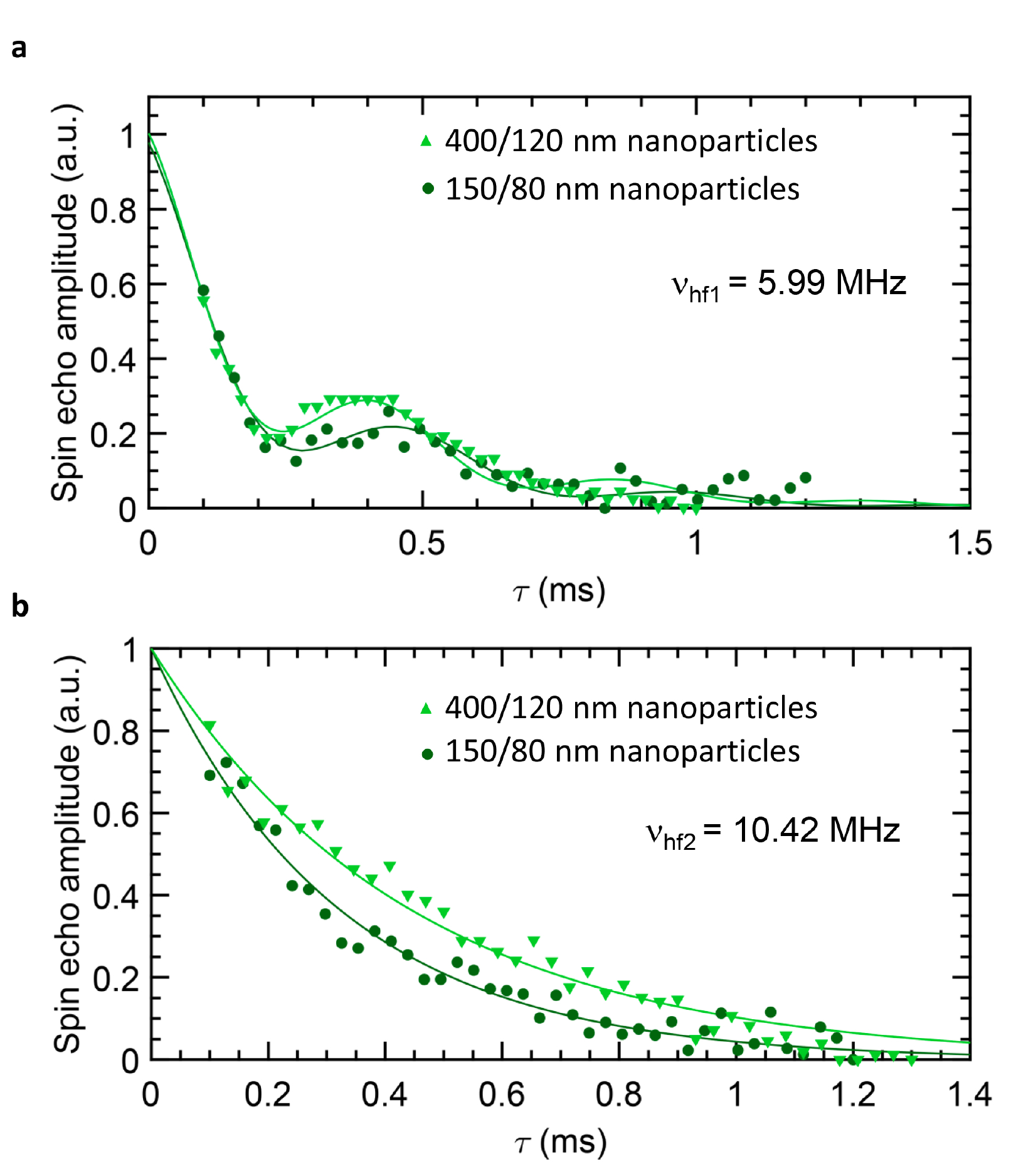}
	\caption{Spin echo decay curves at T = 1.4 K. (a) Spin-echo decay for the 5.99 MHz transition from 400/120 nm diameter particles/crystallites (light green triangles) and 150/80 nm diameter particles/crystallites (dark green dots). Lines correspond to fits by Eq. \ref{eq_fitmodul}. (b) Spin-echo decay for the 10.42 MHz transition for 400 nm (light green triangles) and 150 nm nanoparticles (dark green dots). Lines correspond to single exponential fits.}
	\label{fig.6}
\end{figure}

Spin coherence lifetimes ($T_2$) were finally obtained for both ground-state spin transitions at zero applied magnetic field. The decay of the spin echo amplitude as a function of pulse delay ($\tau$) is shown in Fig. \ref{fig.6} for both sizes of nanoparticles.  Echo envelope modulation is clearly observed for the 5.99 MHz transition ($\pm1/2\leftrightarrow\pm3/2$). This effect has been previously reported in spin transitions of rare-earths as due to the removal of the $\pm M_I$ degeneracy by residual magnetic fields [\onlinecite{arcangeli_spectroscopy_2014}]. Echo envelope modulation has also been attributed to superhyperfine coupling to surrounding nuclear spins [\onlinecite{fravalPrY},\onlinecite{lovric_hyperfine_2011}]. The modulated spin echo decays were fitted to the function [\onlinecite{karlsson}]: 
\begin{equation}
E=Ae^{-2\tau/T_{2}}\hspace{0.1cm}[1+\mathrm{m}\cos^2(\omega\tau/2)] \label{eq_fitmodul}
\end{equation}

where $m$ is the modulation amplitude and $\omega$ the modulation frequency. Spin $T_2$ values of 680$\pm$40 $\mu$s ($\Gamma_h$=470$\pm$50 Hz) and 640$\pm$40 $\mu$s ($\Gamma_h$=500$\pm$50 Hz) were obtained from fit to Eq. \ref{eq_fitmodul} in the 400/120 nm and 150/80 nm diameter particles/crystallites respectively. These values are \DS{longer} than the reported zero-field spin coherence lifetimes in bulk Pr$^{3+}$:\YSO{} [\onlinecite{fraval}]. The modulation frequency, $\omega/2\pi$, was found equal to 2.5 kHz. Unlike Fig. \ref{fig.6}(a), no clear modulation is observed in the $\pm$3/2$\leftrightarrow\pm$5/2 spin echo decays at 10.42 MHz (Fig. \ref{fig.6}(b)). Thus, coherence lifetimes were straightforwardly derived by single exponential fit, yielding spin $T_2$ values of 880$\pm$40 $\mu$s ($\Gamma_h$=360$\pm$40 Hz) for the larger-diameter particles, and 640$\pm$30 $\mu$s ($\Gamma_h$=500$\pm$40 Hz) for the 150/80 nm diameter ones. Spin echo modulations can be expected as long as nuclear Zeeman splittings are comparable to the superhyperfine interaction [\onlinecite{car}]. Since splittings are larger for levels with larger M$_J$, this could explain that modulation is only observed for the $\pm1/2\leftrightarrow\pm3/2$ transition.

\subsection{Discussion}\label{E}

The relaxation and coherence properties here demonstrated for $^{141}$\pr$ $ in nanoscale \YO{} appear very appealing for applications of this material in quantum devices. The results are summarized in Table \ref{table3}. Among them, the long spin dephasing times found in nanoparticles at zero external magnetic field is particularly promising. Unlike the optical transition, spin transitions are rather insensitive to electric perturbations given by their low nuclear Stark coefficient: three orders of magnitude lower than the optical one [\onlinecite{macfarlane_optical_2014},\onlinecite{graf}]. Thus, spin dephasing is here most likely due to magnetic interactions, as concluded for \eu:\YO nanoparticles [\onlinecite{serrano_all-optical_2017}] and in general for many bulk rare-earth doped crystals [\onlinecite{arcangeli_spectroscopy_2014}]. Still, in view of the present results, the magnetic sensitivity of \pr in \YO{} appears lower than in other crystalline hosts including \YSO. This can be explained by the larger crystal field splitting in \YO{} compared to other materials. As mentioned in Sec. \ref{C}, this gives rise to lower second order contributions and results in smaller hyperfine splittings and gyromagnetic ratios (Tables \ref{table1} and \ref{table2}). Furthermore, we note that the spin coherence results presented here are obtained, not from bulk single crystals, but nanoparticles down to 150 nm. Hence, \pr:\YO is a highly performing material at the nanoscale. As a main drawback, the small hyperfine splittings of \pr in \YO{}, especially in the excited state (Fig. \ref{fig.4}) limit the minimum optical pulse length which can be used in quantum memory schemes to about 500 ns. Taking into account the measured optical coherence lifetimes, in the few $\mu$s range (Fig. \ref{fig.2}), shorter pulses would be ideally required to limit dephasing in the excited state in spin-wave quantum memory schemes [\onlinecite{gundogan_solid_2015}]. As a solution, optical $T_2$ enhancement can be envisioned. So far, optical coherence lifetimes in nanoparticles does not appear fundamentally limited [\onlinecite{bartholomew_optical_2017}] and we have already demonstrated that optical $T_2$ extension is obtained by post-synthesis treatments. Further improvement should therefore be possible by increasing the particles crystalline quality and reducing defects [\onlinecite{kunkel}].

\begin{table}
\begin{tabular}{l| c| c c}
\hline  
\hline         
   		 & Ceramic & 400/120 nm & 150/80 nm\\
   		 \hline
   		Optical &      &        & \\
   		$\Gamma_{inh}$ (GHz) & 9 & 27 & \\
   		$T_{1}$ ($\mu$s) & 140 & 140 & \\ 
   		$T_{2}$ ($\mu$s) & 4.5$\pm$0.5 & 3.0$\pm$0.3 & 1.0$\pm$0.1\\
   		$\Gamma_{h,opt}$ (kHz) & 72$\pm$16 &  108$\pm$21 & 315$\pm$64\\
   		       &             &          & \\
   
   		Spin   &             &           & \\
   		$T_{1}$ (s) & 5$\pm$1 & 7$\pm$3  & \\
   		$\Gamma_{inh,5.99}$ (kHz) & 29$\pm$2 & 42$\pm$9 &\\
   		$\Gamma_{inh,10.42}$ (kHz) & 29$\pm$2 & 48$\pm$6 &\\
   		
   		$T_{2,5.99}$ ($\mu$s) & 730$\pm$50 & 680$\pm$40 & 640$\pm$40 \\
   		$\Gamma_{h,5.99}$ (Hz) & 440$\pm$60 & 470$\pm$50 & 500$\pm$50 \\ 
   		$T_{2,10.42}$ ($\mu$s) & 730$\pm$20 & 880$\pm$40 & 640$\pm$30 \\
   		$\Gamma_{h,10.42}$ (Hz) & 430$\pm$30 & 360$\pm$40 & 500$\pm$40 \\     
  \hline
  \hline
  \end{tabular}
  \caption{Summary of spectral and relaxation parameters determined for $^{141}$\pr:\YO ceramic and two sizes of nanoparticles. Spin $T_2$ values for the ceramic sample, not displayed in Fig. \ref{fig.6}, are here given for the sake of comparison.}
\label{table3}
  \end{table}

\begin{figure}
	\centering
	\includegraphics [width=3.5 in]{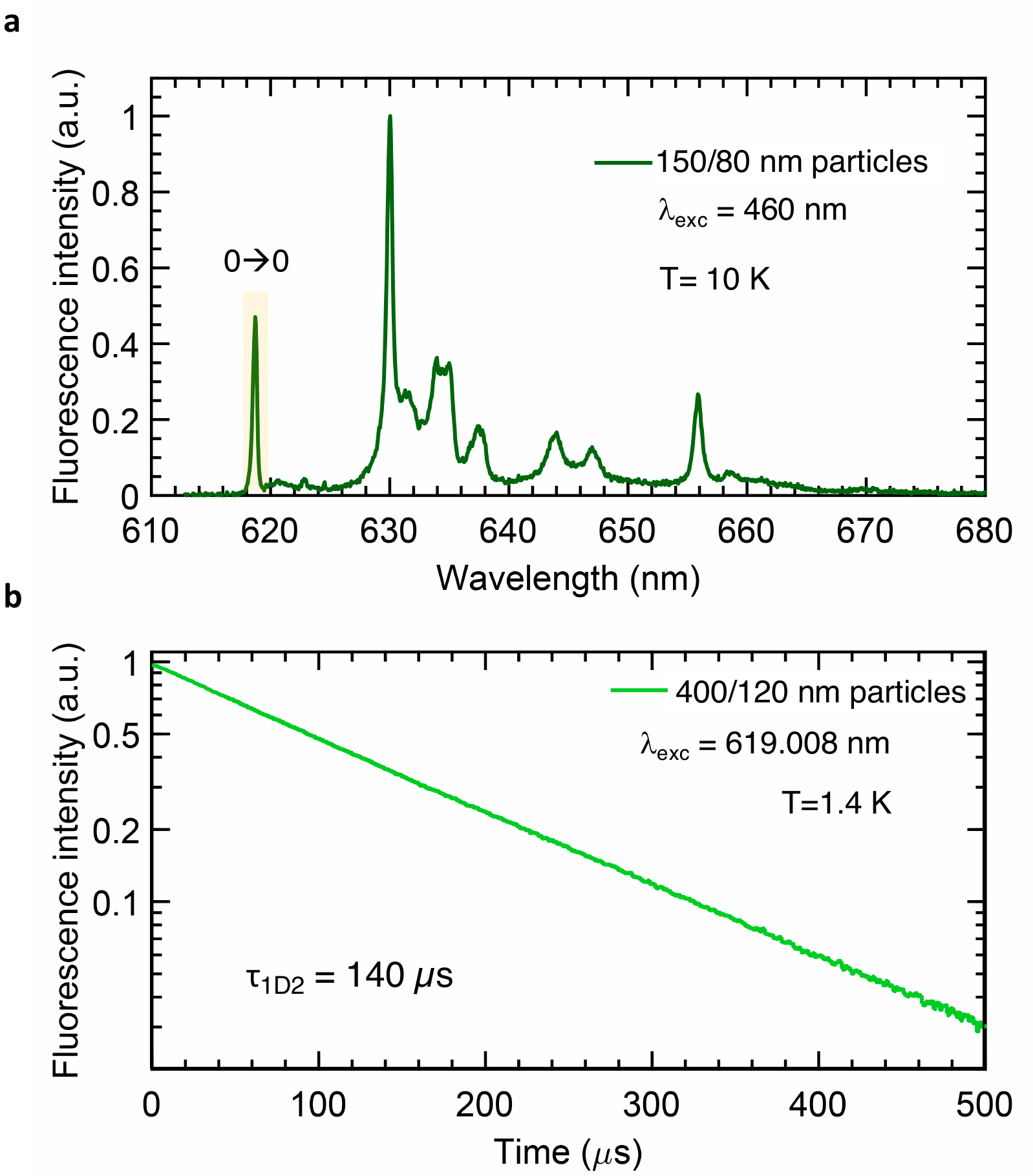}
	\caption{Low temperature fluorescence spectroscopy of \pr:\YO{} nanoparticles. (a) \odt$ $  emissions in the 610-680 nm range. The peaks correspond to transitions towards different crystal field levels in the \thf$ $ ground-state multiplet. The displayed spectrum is corrected from instrumental response.  (b) $^1D_2$ fluorescence decay yielding an optical $T_1$ of 140 $\mu$s.}
	\label{fig.7}
\end{figure}

Apart from quantum storage applications, \pr:\YO nanoparticles coupled to optical micro-cavities appear as promising candidates \DS{for realizing spin photon interfaces emitting} indistinguishable single photons. For this, a \DS{key} parameter to determine is \DS{the effective Purcell factor C= $\xi$$\frac{3\lambda^3}{4\pi^2}$$\frac{Q}{V}$, where $\xi$ is  the branching ratio for the enhanced transition \odt(0)$\rightarrow$\thf(0) (Fig. \ref{fig.7}(a)), $Q$ is the cavity quality factor, $V$ is the mode volume and $\lambda$ is the emission wavelength}. $\xi$ was here derived from the $T_1$/$T_{spon}$ ratio, with $T_{spon}$ the time it would take for the excited state to decay to the ground state through the enhanced transition [\onlinecite{mcauslan}]. To calculate $T_{spon}$ we first determined the transition oscillator strength, $P$, for a single ion, which we found equal to 6.3 10$^{-7}$. We note that in a cubic crystal like \YO, $P$ for a single ion is equivalent to 3 times the average oscillator strength obtained from the absorption spectrum. For $P$=6.3 10$^{-7}$, $T_{spon}$ equals to 2.5 ms, calculated using the expressions in [\onlinecite{mcauslan}]. Thus, with $T_1$=140 $\mu$s (Fig. \ref{fig.7}(b)), we conclude that $\xi$=0.057. For a realistic cavity presenting a finesse $\mathcal{F}$=10$^5$, and smallest mirror separation $d$=2 $\mu$m as the one discussed in [\onlinecite{casabone_cavity-enhanced_2018}], effective Purcell factor C=340 would be expected.  This is more than three times higher than expected for \eu$ $ [\onlinecite{casabone_cavity-enhanced_2018}], mainly because of the difference in branching ratio between \eu$ $ and \pr$ $ transitions. With such Purcell factor, the demonstrated optical T$_2$ values in the 400-nm-diameter and 150-nm-particles (Table \ref{table3}), already verify the condition to emit Fourier transform limited photons: $C>2T_1/T_2$. However, in practice, achieving an effective Purcell factor C=340 implies limiting cavity losses as those due to scattering by the particles. In this sense, the particle size plays an important role with the smaller the particle size the lower the scattering losses. In view of the present results (Table \ref{table3}), efforts to be done go towards the reduction of the the particle size without decreasing the optical $T_2$.

\section{Conclusion}\label{IV}

In the present work we perform a complete high resolution and coherent optical and spin spectroscopic investigation of
$^{141}$Pr$^{3+}$:Y$_2$O$_3$ ceramics and nanoparticles. We report optical $T_2$ values of 3 $\mu$s and 1 $\mu$s for 400 nm and 150-nm-diameter particles respectively. The hyperfine structure of $^{141}$Pr$^{3+}$ in Y$_2$O$_3$ is here for the first time revealed by spectral hole burning and complete
Hamiltonian numerical calculations. Ground-state spin transitions were assigned to 5.99 MHz and 10.42 MHz, for which we demonstrate inhomogeneous linewidths of 42 and 48 kHz. Spin $T_2$ up to 880 $\mu$s was also obtained, a value which exceeds that of bulk Pr$^{3+}$ doped crystals such as Y$_2$SiO$_5$. We finally determined the oscillator strenght and branching ratio for the \pr{}:\odt(0)$\rightarrow$\thf(0) in \YO. These results open up very interesting prospects for nanoscale \pr:\YO{}, in particular as a candidate for quantum information storage and processing and source of indistinguishable single photons. 

\section{Ackowledgements}
This project has received funding from the European Union Horizon 2020 research and innovation program under grant agreement no. 712721 (NanOQTech) and \DS{within the Flagship on Quantum Technologies with grant agreement no. 820391 (SQUARE)}. We thank our colleagues at the MPOE team and collaborators for the transparent ceramic sample used for absorption measurements.

\appendix
\section{}

The  Hamiltonians used in the hyperfine structure calculations  (Sec. \ref{CF}) are given below.
The hyperfine interaction is expressed as:
\begin{equation}
H_{HF}=a_{1}H_1,
\end{equation}		
where 
\begin{equation}
a_{1}=\frac{\mu_{0}}{4\pi}g_{s}\mu_B\mu_{n}g_{n}\left\langle r_{e}
^{-3}\right\rangle. 
\end{equation}
\\
$N$ and $\left\langle r_{e}^{-3}\right\rangle$ are the number and mean inverse cube radius of the $4f$ electrons. Matrix elements of $H_1$, calculated using tensor operator techniques, are given in Ref. [\onlinecite{GuillotNoel:2005bk}]. The quadrupolar interaction is divided in $4f$ electrons ($a_2H_2$) and lattice field ($a_3H_3$) contributions:

\begin{equation}
H_{Q}= a_2H_{2}+a_3H_{3},
\end{equation}
with 
\begin{eqnarray}
a_2 &=&  \frac{-e^2}{4\pi \epsilon_0}(1-R)\left\langle r_{e}^{-3}\right\rangle\frac{Q}{2}, \quad \mathrm{and} \\
a_3 &=&  -\frac{(1-\gamma_\infty)}{(1-\sigma_2)\langle r_{e}^{2}\rangle}\frac{Q}{2}.
\end{eqnarray}

In these expressions, $Q$ is the nuclear quadrupole moment, $\gamma_\infty$, $R$, $\sigma_2$ shielding factors and $\langle r_{e}^{2}\rangle$ the mean square radius of the $4f$ electrons. More details and matrix elements for $H_2$ and $H_3$ can be found in ref. [\onlinecite{guillot-noel_calculation_2010}].

\bibliography{refs} 

\end{document}